\documentclass[showpacs,showkeys,aps,prc,amsmath,amssymb]{revtex4}
\usepackage{exscale}
\usepackage{amsmath}
\usepackage{amssymb}
\usepackage[dvips]{graphicx}
\usepackage{float}
\newcommand{\bea}{\begin{eqnarray}}
\newcommand{\eea}{\end{eqnarray}}
\newcommand{\nn}{\nonumber}

\begin{document}

\title{Short Range Correlations and Spectral Functions in Asymmetric Nuclear Matter}

\author{P. Konrad, H. Lenske, U. Mosel\\
Institut f\"ur Theoretische Physik, Universit\"at Gie\ss en\\
Heinrich-Buff-Ring 16, D-35392 Gie\ss en, Germany}

\begin{abstract}
Dynamical correlations in asymmetric infinite nuclear matter are
investigates in a transport theoretical approach. Self-energies due
to short range correlations and their influence on the nucleon
spectral functions are described in an approach accounting for a
realistic treatment of mean-field dynamics and a self-consistently
derived quasi-particle interaction. Landau-Migdal theory is used to
derived the short range interaction from a phenomenological Skyrme
energy density functional. The spectral functions in asymmetric
nuclear matter are found to follow in their gross features closely
the patterns observed previously in symmetric nuclear matter. An
interesting sensitivity of dynamical self-energies and spectral
functions on the momentum structure of the underlying interactions
is found.
\end{abstract}
\maketitle
\date{\today}

\section{Introduction}\label{sec:Intro}

Correlations beyond those described by the static mean-field are of
genuine interest for the understanding of the dynamics in a nuclear
many-body system. Although experience seems to confirm the
prevalence of mean-field dynamics and the directly related
quasi-particle picture a closer inspection reveals that nucleons in
matter obey more involved rules. For symmetric nuclear matter this
has been discussed extensively in the literature, e.g. in a recent
review \cite{WD2004}, and also was considered in our previous work
\cite{Lehr01,Lehr02,Froemel03}. However investigations for asymmetric
nuclear matter are rare. Some of the few cases are \cite{Bozek2003}
and \cite{hassaneen}. In symmetric matter and stable finite nuclei
one finds typically a suppression of the quasi-particle strength by
about 10\%. Pure mean-field dynamics predicts that in nuclear matter
the spin- and momentum states are uniformly occupied inside the
Fermi sphere up to the sharp surface at the Fermi momentum $k_F$.
Hence, the momentum distribution $n_q(k)$ for protons and neutrons
(q=p,n), respectively, up to a normalization constant is described
appropriately by a step function, $n_q(k)\sim \theta(k^q_F-k)$ where
$k^p_F=k^n_F$ for symmetric nuclear matter. As a matter of fact, in
a more realistic description, going beyond the mean-field, this
pattern changes. The nucleons cease to be in sharp energy-momentum
states but acquire spectral functions of a finite width, thus
resolving the strict on-shell relation between energy and momentum.
Hence, the mean-field configurations only exist with a finite life
time, determined by the strength of the coupling to $2h-1p$ and
$1p-2h$ states, respectively. Correspondingly, in finite nuclei one
observes a reduction of single particle spectroscopic factors when
going away from the Fermi level. More dramatic effects are found in
exotic nuclei. Leaving the valley of stability, the experiments
\cite{Lola2001,Lola2002} and theory \cite{PPNP2001,PPNP2004} agree
in observing strongly suppressed single particle probabilities
already close to the Fermi level of neutron-rich nuclei.

Despite the fact that strongly asymmetric exotic nuclei are of
central interest for modern nuclear structure physics, the
contributions of short range correlations to asymmetric nuclear
matter and finite nuclei is still a rarely touched problem in
nuclear physics. Short range correlations play an important role for
the understanding of nuclear matter as for example seen in the
observed high energy tails of nuclear momentum distributions and
form factors. Also, short range correlations play a decisive role in
understanding the structure of finite nuclei, particle production in
heavy ion collisions and astrophysical objects like supernovae.
Since supernovae are highly asymmetric in isospin we need a more
detailed knowledge of correlations in asymmetric matter beyond
ordinary mean-field interactions.

Here, our main goal is to explore the interplay of mean-field
dynamics and correlations in asymmetric nuclear matter. It is
worthwhile to recall that in asymmetric matter isovector
interactions are giving rise to strongly asymmetric Fermi energies
for proton and neutrons. Hence, as an immediate consequence the
available density of states of excited states changes, becoming
increasingly different for proton and neutron configurations,
respectively, with changing proton-to-neutron ratio. This a new
aspect, clearly not accessible in the conventional approaches
concentrating on symmetric nuclear matter.

From our previous work we know that the spectral functions in
symmetric nuclear matter are determined by interactions of extremely
short range which are very well approximated by point couplings,
reproducing almost perfectly results from rather involved many-body
calculations, e.g. in \cite{benhar}, both for cold
\cite{Lehr01,Lehr02} and hot \cite{Froemel03} symmetric matter at
all densities. It is therefore tempting to extend the same transport
theoretical approach to asymmetric nuclear matter. The fully
microscopic approaches to correlations in asymmetric matter use
either only BHF self-energies from particle-particle ladders in the
T-matrix approach \cite{Bozek2003} or include in addition hole-hole
scattering \cite{deJong96,hassaneen}, corresponding to 2h1p
self-energies but neglecting the 2p1h contributions for particle
states. We go beyond the previous work and consider in this paper
both types of dynamical self-energies, thus obtaining a consistent
description of particle and hole states. However, this is achieved
to the expense on treating the mean-field (BHF) parts
phenomenologically.

A fully self-consistent description as indicated e.g. for symmetric
matter in \cite{deJong96} is beyond the intention of the present
paper. Therefore, long-range mean-field effects are included by a
Skyrme energy functional by which we account for the above mentioned
change in the Fermi energies of protons and neutrons, respectively.
The basic transport theoretical relations are summarized in section
\ref{sec:Transport} and are extended to the general case of
asymmetric nuclear matter. We also discuss the influence of charge
asymmetry on the mean-field, affecting both the effective mass and
the effective binding potentials in asymmetric matter. Landau-Migdal
theory is used to estimate the matrix elements for the $ppp^{-1}$,
$nnn^{-1}$, $pnn^{-1}$ and $npp^{-1}$ collisions by which the
dissipation of spectral single particle strength is determined.
Results of transport calculations for the spectral functions,
self-energies and momentum distributions of protons and neutrons in
asymmetric nuclear matter at various degrees of neutron excess are
presented in section \ref{sec:Results}. Finally we will close in
section \ref{sec:Sum} with a summary and an outlook.

\section{Spectral Functions in Asymmetric Nuclear
Matter}\label{sec:Transport}

The underlying model for our calculations was presented in \cite{Lehr01,
Lehr02} and used for calculations in cold ($T=0$) symmetric nuclear matter at
saturation point density $\rho_0$ and for finte temperature and high
densities in \cite{Froemel03}. At this point we restrict ourself to
a short summary of the formalism  with a few changes in respect to
asymmetric nuclear matter. For a more detailed calculation see
\cite{Lehr01,Lehr02,Froemel03}.

The two-particle-one-hole (2p1h) and one-particle-two-hole (1p2h)
polarization self-energies $\Sigma^>$ and $\Sigma^<$ for nucleons in
nuclear matter are given by \cite{kadanoff}: \bea
    \Sigma^{\gtrless}_q=&&g \sum_{q'}\int \frac{d^3k_2 d\omega_2}{(2\pi)^4}
    \frac{d^3k_3 d\omega_3}{(2\pi)^4}\frac{d^3k_4 d\omega_4}{(2\pi)^4}
    (2\pi)^4\delta^4(k+k_2-k_3-k_4)\overline{|M_{qq'}|^2}\nonumber\\
    &&\times g^{\lessgtr}_{q'}(\omega_2,k_2)g^{\gtrless}_{q}(\omega_3,k_3)
    g^{\gtrless}_{q'}(\omega_4,k_4)
\label{eq:sigma} \eea where $g=2$ is the spin degeneracy. Compared
to symmetric nuclear matter neutrons and protons occupy different
Fermi spheres, which leads to different self-energies. In the
asymmetric case one has to account for collisions with particles of the
same kind (pp and nn) and for collisions with particles of different
kind (pn and np). This explains the summation over the isospin $q'$
in equation (\ref{eq:sigma}).

The one particle correlation function  $g^{\gtrless}_q$ in equation
(\ref{eq:sigma})  can expressed in terms of spectral functions \bea
g^{<}_q(\omega,k)&=& ia_q(\omega,k)\theta(\omega_q-\omega),\nonumber\\
g^{>}_q(\omega,k)&=& -ia_q(\omega,k)(1-\theta(\omega_q-\omega)),
\label{eq:correl} \eea where $\omega_q$ is the Fermi energy of the
nucleons and $\theta(\omega_q-\omega)$ the step function to account
the Fermi distribution in  case of cold nuclear matter. The
nonrelativistic spectral function for nucleons is found explicitly
as, \bea
 a_q(\omega,k)=\frac{\Gamma_q(\omega,k)}
            {(\omega-\frac{k^2}{2 m_q}-\Sigma^{mf}_q-\mathrm{Re}
            \Sigma^{ret}_q(\omega,k))^2+\frac{1}{4}\Gamma^2_q(\omega,k)},
\label{eq:spectral} \eea where $\mathrm{Re}\Sigma^{ret}_q(\omega,k)$
is the real part of the retarded self-energy, which can be
calculated dispersively: \bea
\mathrm{Re}\Sigma^{ret}_q(\omega,k)=P\int\frac{d\omega'}{2\pi}
                                \frac{\Gamma_q(\omega',k)}{\omega'-\omega}
\label{eq:realsigma}
\eea

The width $\Gamma_q(\omega,k)$ is given by the imaginary part of the
retarded self-energy, \bea
    \Gamma_q(\omega,k)=2 \mathrm{Im}\Sigma^{ret}_{q}=
    i(\Sigma^>_q(\omega,k)-\Sigma^<_q(\omega,k)).
\label{eq:gamma} \eea We introduce the asymmetry coefficient \bea
\xi=\frac{\rho_p}{\rho} \eea which indicates the fraction of protons
at a given total isoscalar nucleon density $\rho=\rho_n+\rho_p$. In
a finite nucleus with charge number $Z$ we have $\xi=Z/A$.

To see the influence of asymmetry we split the mean-field in proton
and neutron contributions: \bea
\Sigma_p^{mf}&=&\Sigma_{pp}^{mf}+\Sigma_{pn}^{mf}=\rho \xi u_{pp}
    +\rho(1-\xi)u_{pn}\\
\Sigma_n^{mf}&=&\Sigma_{nn}^{mf}+\Sigma_{np}^{mf}=\rho(1-\xi)u_{nn}
    +\rho \xi u_{np}.
\label{eq:ueff} \eea
$\Sigma_{pp}^{mf}$ and $\Sigma_{np}^{mf}$ are
generated by the proton background, which vanishes for pure neutron
matter. $\Sigma_{nn}^{mf}$ and $\Sigma_{pn}^{mf}$ are generated by
the neutron background. On the right hand side  $u_{pp}$ and
$u_{pn}$ are effective interactions averaged over the proton Fermi
sphere for $u_{pp}$ and the neutron Fermi sphere for  $u_{pn}$ in
phase space. The same for  $u_{np}$ and  $u_{nn}$.

One can easily see in equation (\ref{eq:ueff}) that there is no
difference for neutrons and protons in case of symmetric nuclear
matter ($\xi=0.5$) and one has not to distinguish between protons
and neutrons. Going over to the asymmetric case the mean-field
contributions split up, now protons and neutrons are distinguishable
by their different dynamical laws.  Absorbing the momentum dependent
part of the mean-field up to second order in $k$ into the kinetic
energy term \bea
\hbar^2\frac{k^2}{2m}+\Sigma^{mf}_q(k,\rho)\cong\hbar^2\frac{k^2}{2m_q^{\star}}+U^{eff}_q(\rho)
\eea leads to an effective mass $m_q^{\star}$ and an
momentum-independent effective potential $U^{eff}_q.$ Starting with
symmetric matter, where the effective masses and effective
potentials are the same, for $p$ and $n$, respectively, the masses
and potentials for nucleons of different isospin start to split up
for asymmetric matter. As seen in equation (\ref{eq:spectral}) the
mean-field contribution plays an important role for the spectral
function's pole structure. Hence, we expect a separation of the
proton and neutron quasiparticle peaks.

We account for mean-field effects in our model by means of a Skyrme
energy density functional with parameters taken from the recent set
of ref. \cite{chabanat}. By executing the variation with respect to
the kinetic energy density $\tau_q$ and the nucleon density
$\rho_q$, where $q=n,p$ one obtains for the effective mass and for
the effective potential:
\begin{eqnarray}\label{MeanF}
    \frac{m_q}{m^{\star}_q}&=&1+\frac{2m_q}{\hbar^2}
    (\frac{1}{8}[t_1(2+x_1)+t_2(2+x_2)]\rho\\
    &&+\frac{1}{8}[t_2(2x_2+1)-t_1(2x_1+1)]\rho_q)\nonumber\\
    U^{eff}_q&=&\frac{1}{4}t_0[2(2+x_0)\rho-2(2x_0+1)\rho_q]\nonumber\\
    &&\frac{1}{24}t_3\rho^{\sigma}[2(2+x_3)\rho-2(2x_3+1)\rho_q]\nonumber\\
    &&\frac{1}{24}\sigma t_3\rho^{\sigma-1}[(2+x_3)\rho^2-(2x_3+1)
                          (\rho_p^2+\rho_n^2)\nonumber\\
    &&\frac{1}{8}[t_1(2+x1)+t_2(2+x_2)]\tau\nonumber\\
    &&\frac{1}{8}[t_2(2x_2+1)-t_1(2x_1+1)]\tau_q
\end{eqnarray}
where $\rho$ is the total and $\tau$ total kinetic density. The
Skyrme parameters $x_0$, $x_1$, $x_2$, $x_3$, $t_0$, $t_1$, $t_2$,
$t_3$ and $\sigma$ are given in table \ref{tab:skyrme}.

\begin{table}
    \begin{center}
      \small
        \begin{tabular}{||l|l||} \hline
            \small$t_0$ $(MeV\cdot fm^3)$ & \small$-2490.23$\\
            \small$t_1$ $(MeV\cdot fm^5)$ & \small$489.53$\\
            \small$t_2$ $(MeV\cdot fm^5)$ & \small$-566.58$\\
            \small$t_3$ $(MeV\cdot fm^{3+3\sigma})$ & \small$13803.0$\\
            \small$x_0$ & \small$1.1318$\\
            \small$x_1$ & \small$-0.8426$\\
            \small$x_2$ & \small$-1.0$\\
            \small$x_3$ & \small$-1.9219$\\
            \small$\sigma$ & \small$1/6$\\
            \small$W_0$ $(MeV\cdot fm^5)$& \small$131.0$\\
       \hline
        \end{tabular}
        \normalsize
    \end{center}
    \caption{\label{tab:skyrme}
    The parameters of the Skyrme SLy230a \protect\cite{chabanat} interaction.
    The spin-orbit parameter $W_0$ is included for completeness only.}
\end{table}

In the numerical calculations we use an iterative approach. Starting
with an initial choice for the widths, i.e the imaginary part of the
self-energy, we calculate the spectral functions, which then serve
as input for re-calculating the self-energies. These results are
used as input for the next step of an iterative calculation, thus
improving the results. The calculation is stopped, when the spectral
functions are converged to a given accuracy. In the language of
Feynman diagrams this approach corresponds to the summation of the
sunset diagram to all orders \cite{Lehr02}.

\subsection{The Short Range Interaction}\label{ssec:Interact}

The former investigations were restricted to symmetric nuclear
matter at saturation density. That highly symmetric situation
simplifies the calculations because protons and neutrons are
indistinguishable. Irrespective of the charge state, spectral
functions and other observables can be related to the common Fermi
energy of protons and neutrons. As pointed out before, in asymmetric
matter the isovector mean-field changes the situation completely by
introducing a relative shift of the proton and neutron Fermi
surfaces, in addition to the changes from the differences in
particle number. Also, by following the shift of the equilibrium
density with increasing asymmetry, here we have to have information
on the value of $\overline{|M_{qq'}|^2}$ at changing total number
densities. However, from our former work we expect that only the
short range parts of the full in-medium $NN$ interaction will be
relevant.

An appropriate approach to investigate this aspect and, as an
important side aspect, to test the consistency with the mean-field
part of the model calculations we use Landau-Migdal theory
\cite{migdal}. Given an energy density functional
$\mathcal{E}(\rho)$ Landau-Migdal theory provides a way to calculate
the corresponding residual 2-quasiparticle (2QP) interaction by
second variation of $\mathcal{E}(\rho)$ with respect to the various
spin and isospin densities \cite{migdal,Bender02}. Hence, for an
intrinsically density dependent interaction the corresponding
Landau-Migdal interaction includes various rearrangement terms which
ensure to fulfill a number of important consistency relations
\cite{migdal}. These relations would be violated if the bare
G-matrix interaction would be used instead, as e.g. in
\cite{Bozek2003,hassaneen}.

For a Skyrme energy density functional the resulting parameters
$f_0(\rho)$, $f'_0(\rho)$, $g_0(\rho)$ and $g'_0(\rho)$,
characterizing the interaction strength and density dependence in
the spin-isospin $S=0,T=0$, $S=0,T=1$, $S=1,T=0$, $S=1,T=1$ s-wave
$ph$ interaction channels, are in fact derivable in closed form,
e.g. \cite{Bender02}. According to the structure of the Skyrme
energy density functional the Landau-Migdal parameters are
constrained to monopole and dipole components.

However, these Skyrme values include short and long range
interactions. The origin and nature of the various pieces cannot be
identified directly. Here we are interested mainly in the short
range parts. A simple but meaningful way to extract the short
range parts is to identify the long range components with pion
exchange. Consequently, we define the short range part of the
interaction by subtracting from the interaction derived from the
Skyrme energy density functional the (central part of) the pion
exchange $NN$ interaction. Hence, we use the central interaction
obtained from pion exchange and include anti-symmetrization
explicitly by means of the spin and isospin exchange operators
$P_{\sigma,\tau}$, respectively, thus leading to
\begin{equation}\label{eq:Vpi}
V_\pi(\vec{k}_1,\vec{k}_2)=-f^2_\pi
D_\pi(\vec{k}_1,\vec{k}_2)\vec{\sigma}_1\cdot\vec{\sigma}_2\vec{\tau}_1\cdot\vec{\tau}_2(1-P_\sigma
P_\tau)\quad ,
\end{equation}
which evidently can be arranged into
\begin{equation}\label{eq:VST}
V_\pi(\vec{k}_1,\vec{k}_2)=\sum_{S,T=0,1}{V^{(\pi)}_{ST}(\vec{k}_1,\vec{k}_2)(\vec{\sigma}_1\cdot\vec{\sigma}_2)^S
(\vec{\tau}_1\cdot\vec{\tau}_2)^T}
\end{equation}
Above, the momentum space pion propagator is denoted by
\begin{equation}
D_\pi(\vec{k}_1,\vec{k}_2)=\frac{1}{(p^2+m^2)}F(p^2)
\end{equation}
including a monopole form factor with a cutoff $\Lambda=800~MeV/c$.
The 3-momentum transfer is in the t-channel
$\vec{p}=\vec{k}_2-\vec{k}_1$ while in the u-channel we have
$\vec{p}=\vec{Q}=\vec{k}_2+\vec{k}_1$. Numerically, we use $f_\pi
\simeq 0.075$ which is the standard value for the pseudo-vector $\pi
NN$ coupling constant.

Hence, we cast the energy density functional into the form
\begin{equation}\label{eq:Esk}
E(\rho)=E_s(\rho) + E_\pi(\rho)
\end{equation}
where $E_s\equiv E-E_\pi$ is the short-range contribution. The long
range pionic part $E_\pi$ is given by a sum over the various spin
and isospin transfer contributions as defined in eq.\ref{eq:VST}.
Formally $E_\pi$ can be written as a sum over all spin-isospin
channels
\begin{eqnarray}\label{eq:Epi}
E_\pi(\rho)=
&&\frac{1}{2}\sum_{q,q'=p,n}{\sum_{S,T=0,1}{\int{\frac{d^3k_1}{(2\pi)^3}\int{\frac{d^3k_2}{(2\pi)^3}\Theta(k_F(q)-k_1)
\Theta(k_F(q')-k_2)}}}} \nonumber \\
&&\times V^{(\pi)}_{ST}(\vec{k}_1,\vec{k}_2)
<(\vec{\sigma}_1\cdot\vec{\sigma}_2)^S(\vec{\tau}_1\cdot\vec{\tau}_2)^T>
\end{eqnarray}
although in spin-saturated nuclear matter only the $S=0$ are
non-vanishing. In symmetric, i.e. isospin saturated, nuclear matter
also the $T=1$ components will not contribute. The brackets indicate
traces over spin and expectation values on isospin and the step
functions constrain the momentum integrals to the proton and neutron
Fermi spheres with Fermi momenta $k_F(q)$, $q=p,n$, respectively.

Performing the variation with respect to the various spin and
isospin densities we find in symmetric nuclear matter the set of
Landau-Migdal parameters which in standard notation are
\begin{eqnarray}\label{eq:piLM}
f^{(\pi)}_0(\rho)&=&-4\pi\frac{9}{4} N_0(k_F)\frac{f^2_{\pi}}{2k^2_F} Q_0(1+\frac{m^2_{\pi}}{2k^2_F}) + f^{(r)}(k_F)\\
f'^{(\pi)}_0(\rho)&=&+4\pi\frac{3}{4} N_0(2k_F)\frac{f^2_{\pi}}{2k^2_F} Q_0(1+\frac{m^2_{\pi}}{k^2_F})\\
g^{(\pi)}_0(\rho)&=&+4\pi\frac{3}{4} N_0(k_F)\frac{f^2_{\pi}}{2k^2_F} Q_0(1+\frac{m^2_{\pi}}{2k^2_F})\\
g'_0{}^{(\pi)}(\rho)&=&4\pi N_0(k_F)f_{\pi}^2(\frac{1}{m_{\pi}^2}-\frac{1}{8k^2_F}Q_0(1+\frac{m_{\pi}^2}{2k^2_F}))
\end{eqnarray}
where $Q_0(z)=\frac{1}{2}\ln(\frac{1+z}{1-z})$ is the Legendre polynominal of second kind for $l=0$. A more detailed calculation is found in the appendix. In the
$S=0,T=0$ interaction channel an additional rearrangement term
appears
\begin{eqnarray}\label{eq:Frear}
f^{(r)}(k_F)&=&N_0(k_F)V^{(\pi)}_{0}(0,0)h(k_f/m_\pi)
\end{eqnarray}
where $V^{(\pi)}_{0}(q,k)$ is the monopole component of the
$S=0,T=0$ ph-u-channel pion exchange interaction and
\begin{equation}
h(z) = {\frac { \left( 1+2\,{z}^{2} \right) \ln \left(1+4\,{z}^{2}
\right) -4\,{z}^{2}}{8{z}^{4}}} \quad .
\end{equation}
The Landau-Migdal parameters are pure numbers obtained by
normalizing the interactions to the level density at the Fermi
surface. We use $N_0(k_F)=2k_F M^*/(\pi\hbar)^2$, leading to $1/N_0=
154 MeVfm^3$ at the saturation point with $k_F=1.33 fm^{-1}$.

\begin{figure}[thb]
    \begin{center}
        \includegraphics[width=12cm]{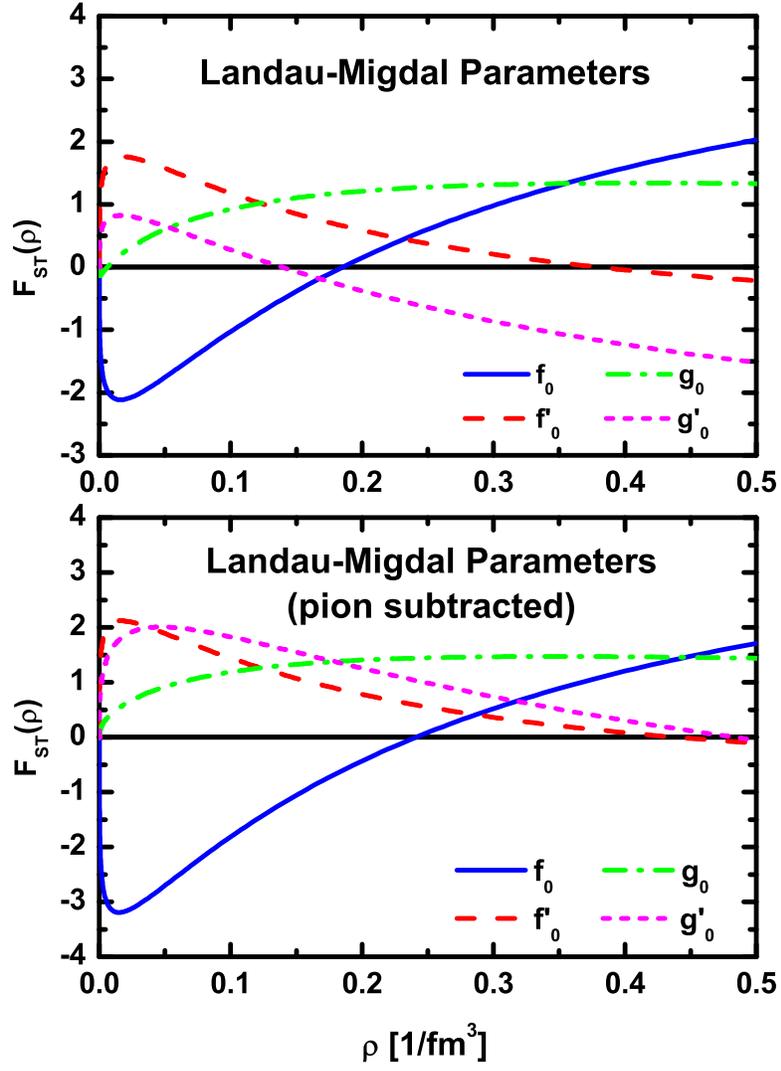}
    \end{center}
    \caption{Landau-Migdal parameters in symmetric nuclear matter for the full Lyon-4 interaction  (upper panel)
    \protect\cite{chabanat} and the pion-subtracted
    short range interaction (lower panel).
    }
    \label{fig:LM}
\end{figure}

The short range components are defined by subtracting the pion
contributions, e.g. in the $(S=0,T=0)$ interaction channel:
\begin{equation}\label{eq:LM_sr}
f^{(s)}_0(\rho)=f_0(\rho)-f^{(\pi)}_0(\rho)
\end{equation}
and  $f'^{(s)}_0$, $g^{(s)}_0$ and $g'^{(s)}_0$ are defined
accordingly. The $\ell=0$ monopole Landau-Migdal Parameters for the
full Lyon-4 interaction and the short range parameters after
subtraction of the pion parts are displayed in Fig.\ref{fig:LM}.
While the pion contributions are small in the $(S=0,T=1)$ and the
$(S=1,T=0)$ channels, the  $(S=0,T=0)$ and $(S=1,T=1)$ gain
additional strength, being attractive in the first and repulsive in
the second case.

\begin{figure}[thb]
    \begin{center}
        \includegraphics[width=15cm]{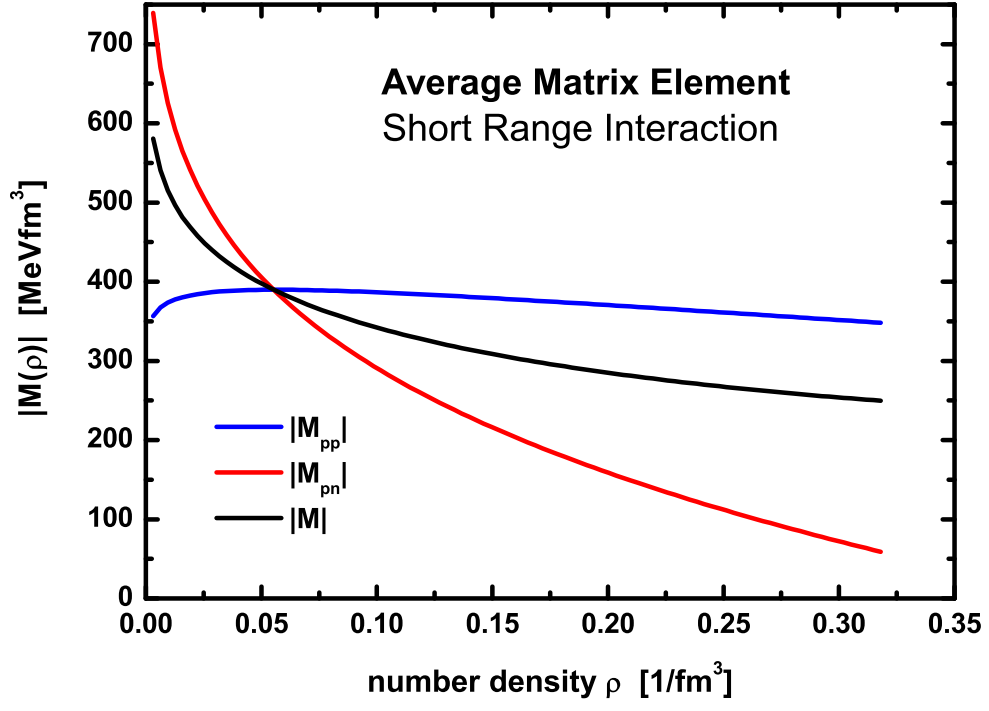}
    \end{center}
    \caption{Dependence of the average short range interaction matrix elements
    $|M_{qq'}$ and of the average matrix element $\overline{M}$,
    eq.\protect\ref{eq:Mbar}, on density in infinite nuclear matter.}
    \label{fig:MSR}
\end{figure}

The short range components are obtained by subtraction of the pion
part from the full Landau-Migdal parameters, e.g. in the $S=0,T=0$
channel: $f^s_0(\rho)=f_0(\rho)-f^\pi_0(\rho)$ and correspondingly
for the other channels. The matrix element relevant for our
calculations is defined from the $pp$, $nn$, $np$ and $pn$
spin-scalar and spin-vector amplitudes, respectively,
\begin{eqnarray}\label{eq:fqq}
f_{q'q}(\rho)&=&f_0(\rho)+(-)^{q'-q}f'_0(\rho);\\
g_{q'q}(\rho)&=&g_0(\rho)+(-)^{q'-q}g'_0(\rho)
\end{eqnarray}
by averaging over the spin degree of freedom. For the isospin
projection quantum numbers we use $q=\pm \frac{1}{2}$ for neutrons
and protons. Because of isospin symmetry we have
$f_{pp}=f_{nn}$ and $f_{pn}=f_{np}$, respectively. Hence, we define
\begin{equation}\label{eq:MSR}
\overline{M}_{q'q}(\rho)=\frac{1}{4}\left(f_{q'q}(\rho)+3g_{q'q}(\rho)\right)
\end{equation}
and averaging out the charge state dependence in a second step,
\begin{equation}\label{eq:Mbar}
\overline{M}(\rho)=\sqrt{\frac{1}{2}(|M_{pp}(\rho)|^2+|M_{pn}(\rho)|^2)}
\quad .
\end{equation}
Results for $\overline{M}_{pp}(\rho)$, $\overline{M}_{pp}(\rho)$ and
$\overline{M}(\rho)$ are shown in Fig.\ref{fig:MSR}. While the short
range interaction among equal particles varies only slowly with
density, the proton-neutron interaction shows a strong density
gradient, reminiscent to the known relations. At saturation density
of symmetric nuclear matter, $\rho=\rho_0=0.16fm^{-3}$ we find
$|\overline{M}(\rho_0)|=341 MeVfm^3$. This value compares extremely
well to the independently derived matrix element
$\widetilde{|M|}=320MeVfm^3$ determined in our previous
investigations \cite{Froemel03,Lehr01,Lehr02} by adjusting the
transport theoretical spectral functions to those from the many-body
calculations of Benhar et al. \cite{benhar}. This very gratifying
agreement confirms the validity of our approach also for asymmetric
matter. Moreover, with the derivation of the relevant interaction
matrix element in a theoretically well justified approach we have
gained predictive power in regions of total density $\rho$ and for
asymmetries $\xi=\rho_p/\rho=Z/A\leq \frac{1}{2}$ which were
inaccessible before.

\section{Correlations in Asymmetric Nuclear Matter}\label{sec:Results}

\subsection{Interactions and Dynamical Self-Energies}\label{ssec:SelfE}

Spectral functions and the imaginary part of the self-energies have
been calculated for nucleons in neutron-rich infinite nuclear matter
in the transport theoretical approach outlined in the previous
sections. Results will be displayed at the saturation density
$\rho_0=0.16 fm^{-3}$ of infinite symmetric nuclear matter although
with decreasing values of the asymmetry $\xi$ the equilibrium point
moves to smaller densities and finally disappears below a critical
value of $\xi$. Our choice of the Lyon-4 parameter set
\cite{chabanat} for the Skyrme energy density functional leads to an
higher effective mass for protons than for neutrons in neutron-rich
matter, i.e. $m^*_p(\rho,\xi)>m^*_n(\rho,\xi)$ for $\xi <
\frac{1}{2}$.

The averaged scattering amplitude was chosen according to
eq.\ref{eq:Mbar}, thus accounting for the variation of the average
matrix element with density. In all calculations, the real part of
the self-energy, which plays an important role for analyticity
\cite{Lehr01}, was included by means of using
eq.(\ref{eq:realsigma}). By numerical reasons the spectral functions
were calculated on energy and momentum grids $(\omega,k)$ with
$-0.5$ $GeV$$\leq\omega\leq +0.5$ $GeV$ and $0\leq k \leq 1.5$
$GeV/c$ which obviously leads to a lack of information on the
behavior of the spectral functions at energies outside the region
covered by the numerical grid. For being able to calculate the
$\mathrm{Re}\Sigma_q$ by eq.(\ref{eq:realsigma}), the imaginary parts were
extrapolated into the regions of large energies by assuming Gaussian
tails.

The spectral properties of nucleons in asymmetric matter are best
explored by keeping the total number density fixed to
$\rho=\rho_{eq}$. Then, the average matrix element is
$(\overline{|M_{qq'}|^2})^{\frac{1}{2}}=350$ $MeV$$fm^3$, being in
agreement both with the derivation sketched above and reproducing
the spectral functions of ref. \cite{benhar}.

\begin{figure}
    \begin{center}
        \includegraphics[scale=0.60]{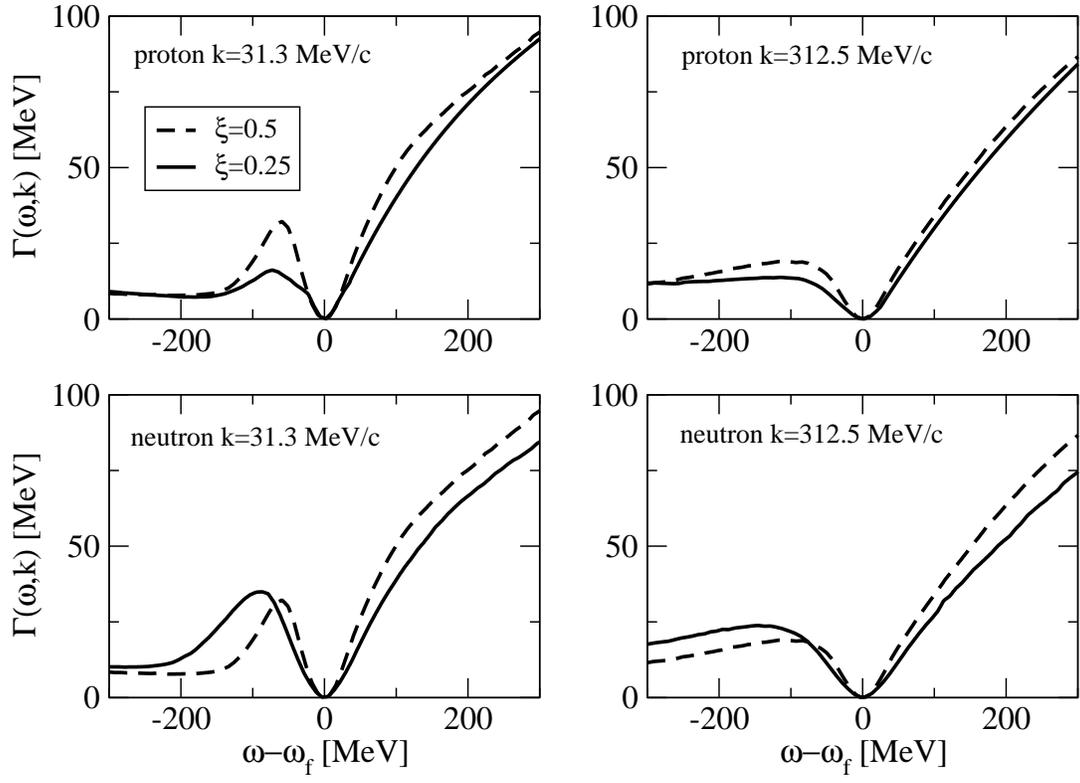}
    \end{center}
    \caption{The nucleon width for symmetric (dashed line) and asymmetric nuclear matter (full line). The upper graphs refer to protons the lower ones to neutrons.}
    \label{fig:sigasymver}
\end{figure}

\begin{figure}
    \begin{center}
        \includegraphics[scale=0.50]{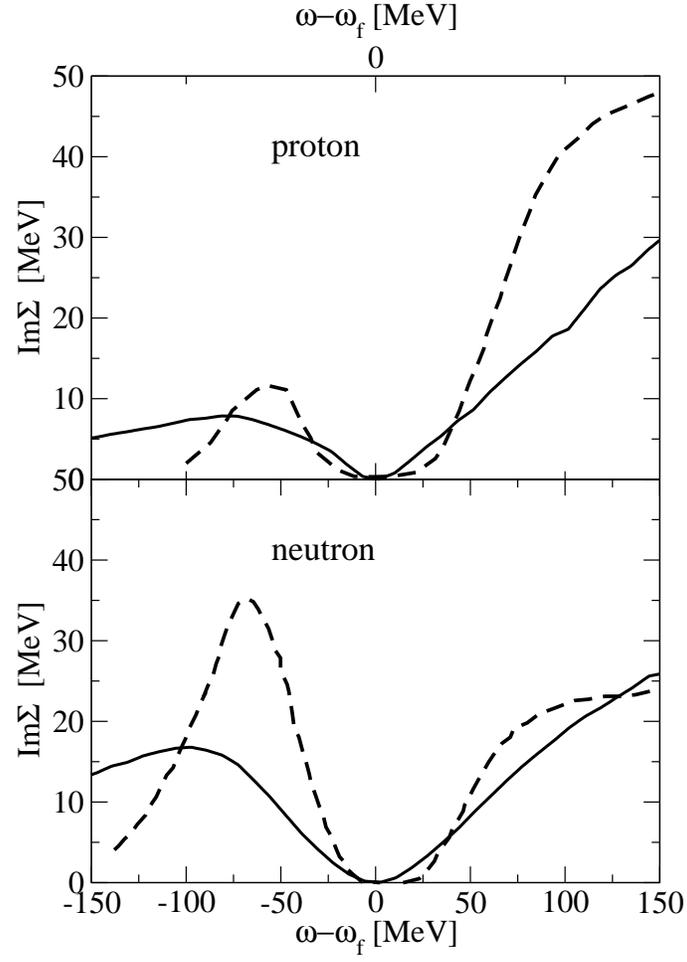}
    \end{center}
    \caption{The imaginary part of the nucleon self-energy for momentum $k=0.4 k_f$ and asymmetry $\xi=0.25$ at a density $\rho=0.17$ $1/fm$.
    The dotted line represents the BHF results from \cite{hassaneen}.}
    \label{fig:tuebver}
\end{figure}

In the imaginary part of the self-energies dynamical correlations
are reflected most clearly and directly. Results for
$\mathrm{Im}\Sigma_q\equiv \frac{1}{2}\Gamma_q$ are displayed in
fig.\ref{fig:sigasymver} for protons and neutrons (q=p,n),
respectively, at an asymmetry $\xi=\frac{1}{4}$. The overall
dependence of $\mathrm{Im}\Sigma_q$ on energy is constrained by the
requirement that the ground state is stationary. Theoretically, this
is taken care off by using retarded Green functions, leading to
$\mathrm{Im}\Sigma_q(\omega_F(q))\equiv 0$ as seen in
fig.\ref{fig:tuebver} and guaranteeing that the ground state is
indeed stable as indicated by the vanishing width.

In Fig.\ref{fig:tuebver} also results from a recent Brueckner
Hartree-Fock (BHF) study of the T\"ubingen group \cite{hassaneen}
are included. Both approaches agree qualitatively in the global
energy dependence given by a strong increase of
$\mathrm{Im}\Sigma_q$ at energies away from the Fermi surface. But
in detail, differences are seen to emerge on the quantitative level
and especially in the hole sector: While our approach predicts long
tails of $\mathrm{Im}\Sigma_q$ extending to large values of
$|\omega-\omega_F|$ and eventually levelling off by slowly
approaching asymptotic values, the BHF calculations show a much more
pronounced fall off at energies deep in the Fermi sea. Such a cutoff
behavior is somewhat unexpected at first sight because the density
of intermediate $(2h1p)$ states still increases with increasing
energy which should overcompensate the suppression introduced by the
momentum structure of the interactions. The deviations in the
particle sector ($\omega > \omega_F$) are reflecting the differences
in the interactions. The BHF calculations are based on the CD-Bonn
NN-potential which seems to provide a larger amount of hard
scattering at large off-shell momenta than our simplified
description using a zero-range interaction, obviously not
distinguishing between on- and off-shell processes. However, the
results of ref.\cite{hassaneen} show that similar differences are
observed in the BHF calculations when using different NN-potentials.
This points to an interesting - and rare - sensitivity of the
dynamical self-energies on the off-shell momentum structure of
interactions.

The striking similarity of our results to the full BHF calculations
is a strong argument for the universality of the underlying
dynamical effects. At first sight it might be surprising that the
close relationship is obtained despite the fact that our approach
does not include explicitly the tensor interaction. Tensor
interactions play a crucial role in BHF and also relativistic
Dirac-BHF (DBHF) calculations \cite{deJong96}. The tensor
interaction is of rank-2 in the spin and the orbital variables and
as such has a vanishing expectation value for the ground state of
spin saturated nuclear matter which we consider here. However, the
tensor interaction contributes from second order on to the equation
of state of symmetric and asymmetric nuclear matter. In the BHF and
also the DBHF calculations the tensor parts introduce an especially
strong proton-neutron correlation in the particle-particle channel
\cite{Bozek2003,deJong96,hassaneen}. Obviously, the on-shell part of
that effect is accounted for by the phenomenological Skyrme
functional but we cannot exclude that off-shell contributions might
be missing.

A slightly different situation is encountered in the dynamical
self-energies. As seen from eq.\ref{eq:sigma}, these self-energies
are of at least second order in the residual interaction leading
from the ground state to the intermediate excited states and back to
the ground state. In these processes, the tensor interaction
contributes in principle through the excitation of spin and
spin-isospin modes in the particle-hole channel. However, since the
corresponding particle-hole interactions are repulsive the spectral
strengths of these modes are shifted to higher energies. Also, as
seen in Fig.\ref{fig:LM} the interactions in the spin-isospin
channels are relatively weak, inhibiting collective enhancements as
observed in the spin-independent channels. Hence, the spectral
properties close to the Fermi-surface are only weakly affected
although contributions at higher energies and in the far off-shell
regions cannot be excluded. Still, the close agreement with the BHF
results of ref.\cite{hassaneen} indicates that such effects cannot
play a major role.

In free space the tensor interaction plays a crucial role in the
proton-neutron system as reflected by the binding of the deuteron.
In the present context one could expect a correlation in the
particle-particle channel. However, explicit calculations by other
groups have shown that such correlations die out quickly with
increasing density. After all, the bulk properties of nuclei are
well understood by nucleonic degrees of freedom without the need to
introduce a significant amount neither of condensed deuteron matter
nor $pn$ pairing in the isospin $I=0$ channel, to the best of our
knowledge. The reason is that the in-medium tensor interactions and,
accordingly, the deuteron-like particle-particle channels are found
to be suppressed with increasing densities \cite{Roepke00,Zamick95}.
For example, the deuteron as a bound state disappears beyond the
so-called Mott density \cite{Roepke82,Schmidt90}. But as a reminder
a pronounced $pn$ correlation might survive at higher densities. As
a side remark we note that the situation is different when
investigating processes especially magnifying the in-medium pair
channels like photo-absorption on nuclei above the giant dipole
resonance where pair processes play a significant role (see e.g.
\cite{Benhar03}).

\subsection{Proton and Neutron Spectral Functions}\label{ssec:SpecF}

\begin{figure}
    \begin{center}
        \includegraphics[scale=0.60]{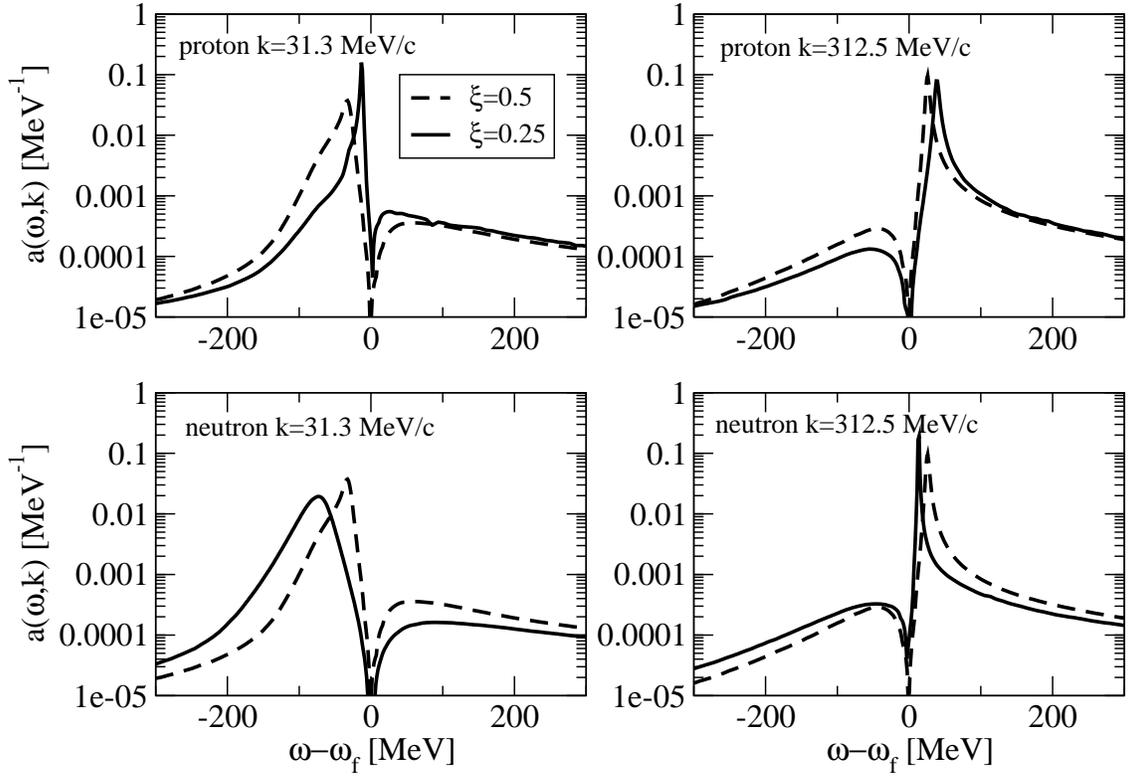}
    \end{center}
    \caption{The nucleon spectral function for symmetric and asymmetric nuclear matter. The dashed line refers to symmetric and the straight line to asymmetric nuclear matter with an asymmetry of $\xi=0.25$. The upper graphs show the results for protons, the lower ones the results for neutrons.  }
    \label{fig:spekasymver}
\end{figure}

In Fig. \ref{fig:spekasymver} proton and neutron spectral functions
at $\rho=\rho_{eq}$ are displayed for two different momenta and
asymmetries $\xi=\frac{1}{2},\frac{1}{4}$. In all cases, the
quasiparticle peaks are clearly identified, as one would expect from
symmetric nuclear matter. The main difference to symmetric nuclear
matter are in fact the changes in the mean-field due to the
isovector self-energy which translate into corresponding changes in
the chemical potentials. With increasing neutron excess the protons
are bound increasingly stronger while the effective neutron
potential becomes shallower, leading to less binding. Thus, the
number asymmetry induces via the isovector interactions a gap
between the proton and neutron Fermi surfaces. Hence, both the shape
and the location of the peak are influenced by the isovector
interactions through their influence on the chemical potentials.

Comparing the results for $\xi=\frac{1}{4}$ with those for symmetric
matter, $\xi=\frac{1}{2}$, we find from Fig.\ref{fig:spekasymver}
that in asymmetric matter the proton hole strength functions are
moved towards the Fermi level while the proton particle strength is
moved away from the Fermi level. The neutron spectral distributions
show the complementary behavior: the hole strength is shifted down
into the Fermi sea while the particle strength is approaching the
Fermi level. In neutron-rich matter the Fermi momentum for protons
is smaller than in symmetric nuclear matter and for neutrons vice
versa. We first consider the spectral functions in the hole sector,
i.e. for states in the Fermi sphere $k_q\leq k_{Fq}$. The hole
spectral functions peak are close the mean-field on-shell energy
position \bea \omega^q_{on}=\frac{\hbar
k^2}{2m_q^{\star}}+U^{eff}_q-\omega_{fq}=\frac{\hbar
(k^2-k_{Fq}^2)}{2m_q^{\star}}. \eea For a momentum of $k=31.3$
$MeV/c$ this leads to a proton quasiparticle peak at about
$\omega^p_{on}=-28$ $MeV$, while the neutron quasiparticle position
is at $\omega^n_{on}=-78$ $MeV$, as seen in
Fig.\ref{fig:spekasymver}. Hence, the proton peak is clearly shifted
towards the Fermi edge, while the neutron strength is repelled.
Similar observations are made in the particle sector. For $k=312.5$
$MeV/c$, corresponding to $\omega (p)=35$ $MeV$ and $\omega (n)=8$
$MeV$ respectively, the quasiparticle peak for the neutron is
shifted towards the Fermi level, while the quasiparticle peak for
the neutron is shifted away.

Another observation from Fig.\ref{fig:spekasymver} is that the width
of the quasi-particle peaks decreases when approaching the Fermi
surface. This is in agreement with the general expectation that
closer to the Fermi surface the life time of quasi-particle states
increases strongly. The energy-momentum structure of
$\Gamma(\omega,k)$ is illustrated in Fig.\ref{fig:sigasymver} for
two momentum cuts in symmetric and asymmetric matter. From the
behavior of $\Gamma(\omega,k)$ we conclude that in asymmetric matter
the collision rates in the hole sector increase for neutrons while
for protons they slightly decreases. For either type of nucleon the
amount of correlations decrease for states above the Fermi level
compared to symmetric matter. These results indicate that the real
and imaginary parts of the polarization self-energies roughly scale
with the density of states of the same kind of particles. Similar
observation have been made in \cite{hassaneen}.

\begin{figure}[thb]
    \begin{center}
        \includegraphics[width=10cm]{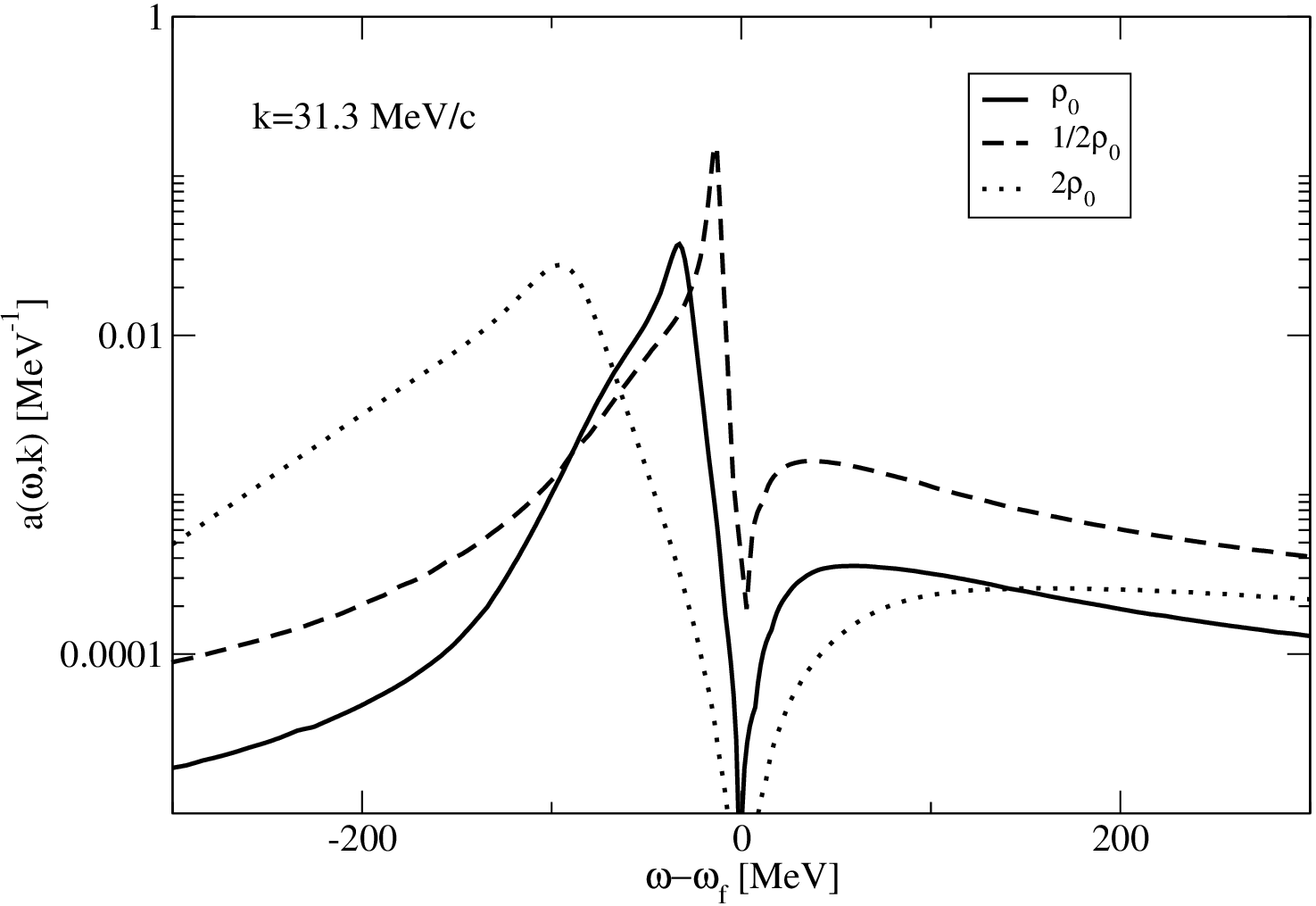}
    \end{center}
    \caption{Spectral functions of a nucleon well inside the Fermi sea at density
    $\rho=\frac{1}{2}\rho_{eq}$, $\rho=\rho_{eq}$ and $\rho=2\rho_{eq}$.}
    \label{fig:spec1}
\end{figure}

\begin{figure}[thb]
    \begin{center}
        \includegraphics[width=10cm]{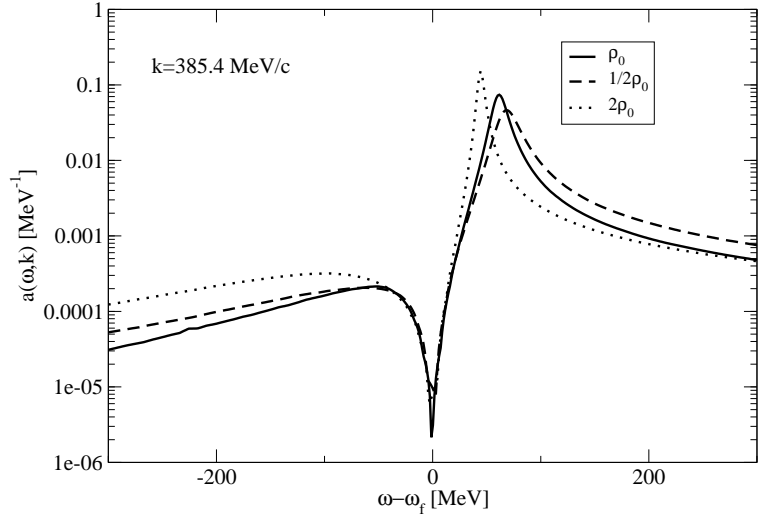}
    \end{center}
    \caption{Spectral functions of a nucleon well outside the Fermi sea at density
    $\rho=\frac{1}{2}\rho_{eq}$, $\rho=\rho_{eq}$ and $\rho=2\rho_{eq}$.}
    \label{fig:spec2}
\end{figure}

The dependence of spectral functions in nuclear matter on density
is investigated in Fig.\ref{fig:spec1} and Fig.\ref{fig:spec2}.
Spectral functions for symmetric matter and at total
densities between half and twice nuclear saturation density
$\rho_{eq}=0.16fm^{-3}$ are displayed. The results are obtained by using the density dependent matrix element from equation \ref{eq:Mbar}.  
In Fig.\ref{fig:spec1} the spectral distributions of a nucleon well inside the Fermi sea are
displayed. The evolution of the spectral functions with density
reflect the changes in the chemical potentials. At
$\rho=\frac{1}{2}\rho_{eq}$ a state with the chosen momentum is much
closer to the Fermi surface than for $\rho=\rho_{eq}$ or
$\rho=2\rho_{eq}$. This explains the variation in width and position
of the quasi-particle peak seen in Fig.\ref{fig:spec1}, seen to be
shifted into the Fermi sea with a considerable gain of width and
high energy strength when the density increases.

In Fig.\ref{fig:spec2} spectral functions for a particle state well
above the Fermi surface are shown. Here, a complementary behavior is
found. With increasing density the relative distance of the
quasi-particle from the Fermi surface decreases which is reflected
in a reduction of the energy shift and the width of the
quasi-particle peak at higher density. In both cases, long spectral
tails in the region below or above the Fermi surface, respectively,
are found, indicating a considerable shift of single particle
strength away from the quasi-particle peak. The overall features,
however, are qualitatively similar to those observed previously in
symmetric nuclear matter \cite{Lehr01,Lehr02} although the results
differ on the quantitative level.

\subsection{Proton and Neutron Momentum
Distribution}\label{sec:MomDis}

\begin{figure}
        \begin{center}
                \includegraphics[scale=0.60]{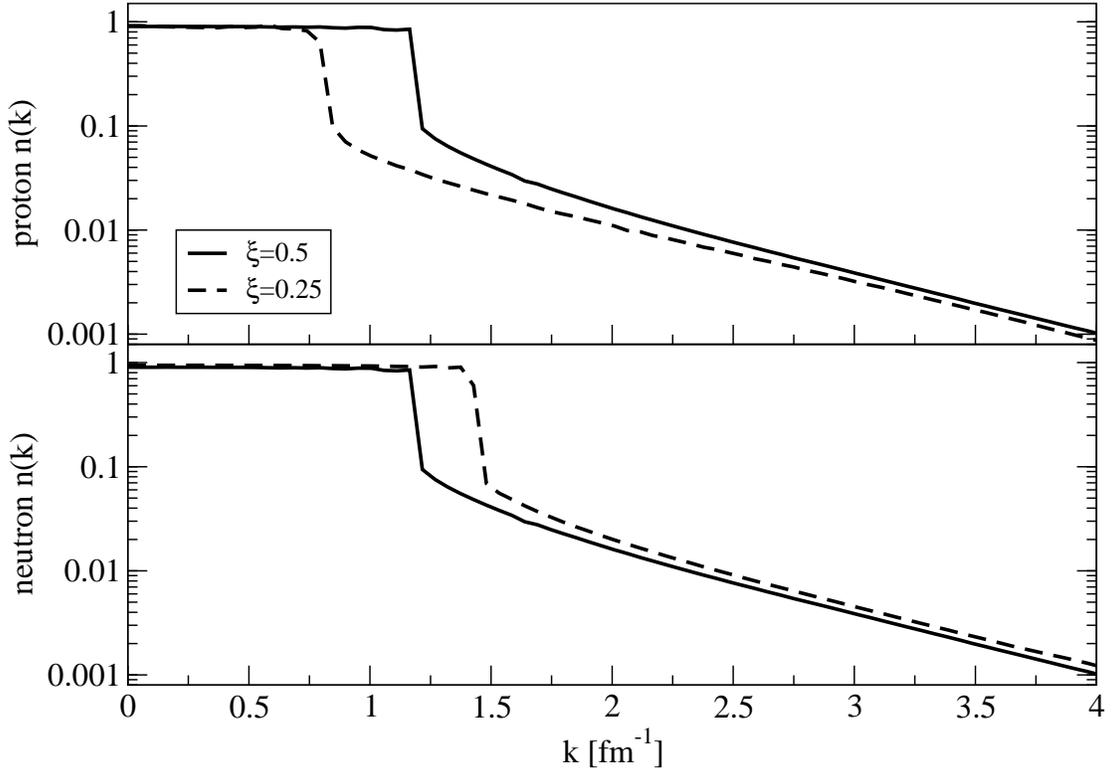}
        \end{center}
        \caption{The momentum distribution for protons (upper graph) and neutrons (lower graph). The straight line is the result for symmetric nuclear matter the dashed line for asymmetric nuclear matter ($\xi=0.25$)}
        \label{fig:momdis}
\end{figure}

We finally consider the proton and neutron one-body ground state
momentum distributions
\begin{equation}
n_q(k)=\int^{\mu_q}_{-\infty}{d\omega a_q(\omega,k)}
\end{equation}\label{eq:mdis}
where $\mu_q$ is the chemical potential for nucleons of type
$q=p,n$. Results for symmetric and asymmetric nuclear matter are
shown in Fig.\ref{fig:momdis}. There, the shrinking of the proton
Fermi sphere in neutron-rich matter is clearly visible in terms of
the diminished region of occupied proton momentum states. In
asymmetric matter the structure of the momentum distributions
follows otherwise closely the same pattern as found already in
symmetric matter: Inside the Fermi sea the occupancy is reduced by
about 10\% and this fraction of the spectral strength is shifted
into a high momentum tail with an almost exponential decline. At
first sight it seem surprising to find that at high momenta the
distributions approach the each other showing very similar
magnitudes and slopes. This result is another illustration of the
universality of the dynamical short range correlations which already
was found in our previous investigations
\cite{Froemel03,Lehr01,Lehr02}. The bulk parts of proton and neutron
matter are seen to be contained in regions $k\leq k_F(q)$, where
obviously, $k_F(p)\leq k_F(n)$ in neutron-rich matter. The short
range correlations produce a similar pattern  as in symmetric matter
by depopulating the Fermi sea by about $10\%$ irrespective of the
asymmetry. In fact in all cases the high momentum tails are almost
identical, thus confirming the universality of the underlying
dynamical processes.

\begin{figure}
        \begin{center}
                \includegraphics{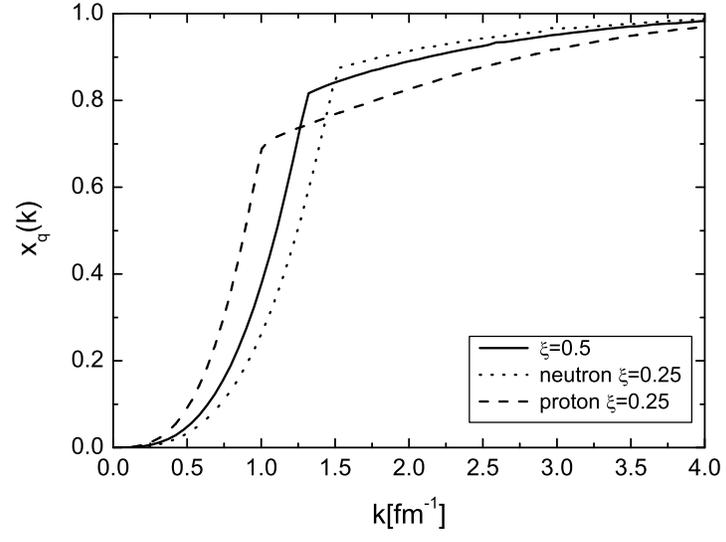}
        \end{center}
        \caption{The the cumulative density index for symmetric nuclear matter (full line) and for protons (dashed line) and neutrons (dotted line) in neutron-rich matter with $\xi=0.25$.}
        \label{fig:occupation}
\end{figure}

Integrating the proton and neutron momentum-distributions over a
sphere with radius $k$ in  momentum-space and normalizing the result
to the density we obtain the cumulative density index \bea
x_q(k)=\frac{1}{\rho_q \pi^2}\int_0^k dk'n_q(k')k'^2 \label{eq:rhok}
\eea where $x_q(k)\rightarrow 1$ for $k\rightarrow\infty$. The
quantity $x_q$ is particular well suited to display the fractional
relative exhaustion of the given density up to a finite momentum
$k$. For a system of non-interacting quasi-particles we expect
$x_q(k) \sim (\frac{k}{k_F})^3\theta(k_F-k)+\theta(k-k_F)$. In a
system with correlations the Fermi surface will be visible as a
discontinuity in the slope as seen in Fig.\ref{fig:occupation}. The
results displayed in Fig.\ref{fig:occupation} show very directly the
significant difference in the correlations pattern of protons and
neutrons: A large fraction of the proton strength is shifted into
the high momentum region as indicated by the slow convergence of
$x_p(k)$ towards unity. For symmetric nuclear matter we find from
equation \ref{eq:rhok} that about $82\%$ of the states in the Fermi
sphere ($k\leq k_{Fq}$) are occupied, while in neutron-rich matter
with $\xi=\frac{1}{4}$ only $71\%$ of the proton but $87\%$ of the
neutron states inside the Fermi sphere are occupied. This
observation is on a qualitative level in agreement with BHF
calculations \cite{hassaneen}. Interestingly, we have obtained the
asymmetric depletion of proton and neutron occupation without using
explicitly a tensor interaction.

\section{Summary and Outlook}\label{sec:Sum}

Dynamical correlations in asymmetric infinite nuclear matter were
investigated in a transport theoretical approach, thereby extending
our previous work into a new regime. It is worthwhile to emphasize
that only few studies on dynamical correlations in asymmetric exist.
In asymmetric matter isovector effects are strongly enhanced with
increasing asymmetry. On the mean-field level this was taken into
account by using a modern Skyrme energy density functional from the
Lyon group \cite{chabanat} with parameters adjusted both to finite
nuclei and neutron stars. In this sense, we describe the evolution
of mean-field dynamics with asymmetry and density in a realistic
model. In particular, the saturation properties of infinite nuclear
matter are well described. Using Landau-Migdal theory we could
derive from the Skyrme functional the corresponding effective
quasi-particle interaction in a self-consistent way. An approach was
presented which allows to extract the short-range interactions by
subtracting the pion exchange contributions.

The overall features of self-energies and spectral functions in
asymmetric nuclear matter resemble those found in our previous work
on short range correlations in symmetric matter. In particular, the
present investigation confirms our previous conjecture on the
universal character of short range correlations in infinite nuclear
matter. An interesting observation in the present context is the
sensitivity of the spectral distributions on the momentum structure
of self-energies and interactions.

\section*{Acknowledgments}
We thank F. Fr\"omel for helpful discussions and numerical hints during preparation of this work.

\begin{appendix}
\section{Landau-Migdal parameters for the pion}
Second variation of the energy density in equation (\ref{eq:Epi}) with respect to spin and  isospin densities leads to
\bea
\frac{\partial^2 E_{\pi}(\rho)}{\partial \rho_q \partial \rho_{q'}}&=& \sum_{S,T=0,1}[V_{ST}(k_f(q),k_f(q'))(\vec{\sigma}_1\cdot\vec{\sigma}_2)^S(\vec{\tau}_1\cdot\vec{\tau}_2)^T\\
&+& \frac{1}{k_f(q')^2}\sum_{q''}\int dk k^2 \Theta(k_f(q'')-k)\frac{\partial}{\partial q}_{|k_f(q')}V_{ST}<(\vec{\sigma}_1\cdot\vec{\sigma}_2)^S(\vec{\tau}_1\cdot\vec{\tau}_2)^T>]\nn
\label{eq:2ndEpi}
\eea
where the brackets indicate ground state expectation values.
From which we now identify the Landau-Migdal parameters:
\bea
f^{(\pi)}&=& N_0(k_f)V_{00}+f^{r}(k_f)\\
f'^{(\pi)}&=& N_0(k_f)V_{10}\\
g^{(\pi)}&=& N_0(k_f)V_{01}\\
g'^{(\pi)}&=& N_0(k_f)V_{11}
\eea
\end{appendix}
where $f^{r}(k_f)$ is an rerrangement term, which appears for spin-saturated symetric nuclear matter only in the $S=0$, $T=0$ channel. Expansion of the in-medium scattering amplitude in a Legendre series, which is common, leads to
\bea
f^{(\pi)}_l=\frac{1}{2}\int_{-1}^{1} P_l(\cos \theta) f^{\pi}(\cos \theta) d\cos\theta.
\eea
The integration over the angle $\theta$ is done using the Newman formula
\bea
Q_l(t)=\frac{1}{2}\int_{-1}^1\frac{P_l(x)}{t-x}dx
\label{eq:newman}
\eea
where the $Q_l(t)$ are Lendendre polynominals of the 2nd kind. Using equation (\ref{eq:newman}) together with our expressions for the Landau-Midgal parameters leads to
\bea
f_l^{(\pi)}&=&-4\pi\frac{9}{4} N_0(k_F)\frac{f_{\pi}^2}{2k^2_F}Q_l(1+\frac{m_{\pi}^2}{2k^2_F})+f_l^{(r)}(k_F)
\label{eq:fl}
\\
f'_l{}^{(\pi)}&=& 4\pi\frac{3}{4} N_0(k_F)\frac{f_{\pi}^2}{2k^2_F}Q_l(1+\frac{m_{\pi}^2}{2k^2_F})\\
g_l^{(\pi)}&=& 4\pi\frac{3}{4} N_0(k_F)\frac{f_{\pi}^2}{2k^2_F}Q_l(1+\frac{m_{\pi}^2}{2k^2_F})\\
g'_l{}^{(\pi)}&=&g'_H\delta_{l0}-4\pi\frac{1}{4} N_0(k_F)\frac{f_{\pi}^2}{2k^2_F}Q_l(1+\frac{m_{\pi}^2}{2k^2_F}).
\eea
where the $l=0$ components in the $S=1$, $T=1$ channel includes in addition the direct (i.e. Hartree-type) contributions
\bea
g'_H&=&4\pi N_0(k_F)\frac{f_{\pi}^2}{m_{\pi}^2}
\eea
For the additional rearrangement term in equation ($\ref{eq:fl}$) we get
\bea
f^{(r)}(k_F)&=&N_0(k_F)\frac{1}{k^2_F}\int{dkk^2\Theta(k_F-k)\frac{\partial}{\partial
q}_{|k_F} V^{(\pi)}_{0}(k,k')}
\eea
where $V_0^{\pi}(k,k')$ is the monopol component of the $S=0$, $T=0$ u-channel pion exchange interaction.

\newpage

\begin{figure}[thb]
    \begin{center}
        \includegraphics[width=15cm]{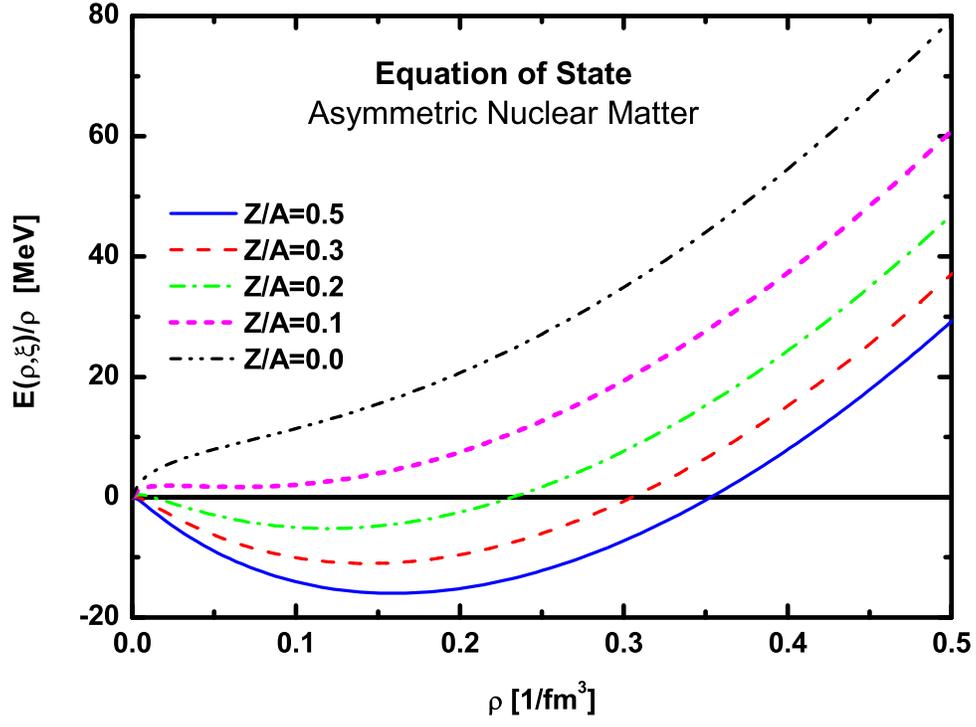}
    \end{center}
    \caption{Equation of state of infinite nuclear matter at various asymmetries $\xi=\rho_p/\rho=Z/A$.
    Results obtained with the Lyon-4 parameter set \protect\cite{chabanat} are shown.
    }
    \label{fig:EOS}
\end{figure}


\end{document}